\documentstyle[amsfonts,epsfig,preprint,aps]{revtex}
\tightenlines
\begin{document}
\draft
\preprint{ }

\title{A self-consistent Ornstein-Zernike approximation for the Random Field Ising model.}
\author{E. Kierlik, M. L. Rosinberg, and G.Tarjus}
\address{Laboratoire de Physique Th\'eorique des Liquides \cite{AAAuth},
Universit\'e Pierre et Marie Curie,\\ 
4 Place Jussieu, 75252 Paris Cedex 05, France}
\maketitle

\begin{abstract}
We extend  the self-consistent
Ornstein-Zernike approximation (SCOZA), first formulated in the context of liquid-state theory, to the study of the random field Ising model.
 Within the replica formalism, 
we treat the quenched random
field as an annealed spin variable, thereby avoiding the usual average over the random field distribution. This allows to study the
influence of the distribution on the phase diagram in finite dimensions. The thermodynamics and the correlation functions are obtained
as solutions of a set a coupled partial differential equations with magnetization, temperature and disorder strength as independent variables.
A preliminary analysis based on high-temperature and $1/d$ series expansions shows that the theory can predict accurately the dependence of the
critical temperature on disorder strength for dimensions $d>4$. For the bimodal distribution, we find a tricritical point which moves to weaker fields 
as the dimension is reduced. For the Gaussian distribution, a tricritical point may appear for $d$ slightly above $4$.
\end{abstract}
\pacs{{\bf Key words}: Disordered systems, Ornstein-Zernike equations, Random Field Ising model.}
\newpage

\def\be{\begin{equation}}
\def\ee{\end{equation}}
\def\bea{\begin{eqnarray}}
\def\eea{\end{eqnarray}}
\def\bit{\begin{itemize}}
\def\eit{\end{itemize}}

\section{Introduction}

Our understanding of the full phase diagram of randomly disordered magnetic systems in finite dimensions 
is severely hampered by the lack of a theory that takes into account the effect of fluctuations in an approximate but sensible fashion.
This is probably the reason why the nature of the paramagnetic to ferromagnetic transition in the random field Ising model (RFIM) is 
still under debate after nearly twenty years of intensive studies (for reviews see \cite{NV88,BY91,N97}). The influence of the random field 
distribution on the order of the transition is one of the important open problems. Whereas mean-field theory \cite{SP77,A78},
which should be valid in high dimensions, predicts that some distributions give rise to a tricritical point,
neither numerical simulations \cite{YN85,OH86,RY93,R95} nor high-temperature series expansions \cite{SA82,H85,GAAHS93} 
have been able to yield a clear-cut answer yet. Recent numerical determinations of the ground states at zero temperature do not seem to clarify
this issue \cite{SBMCB97,AS97}. In addition, as suggested by several analytical works \cite{AB87,MY92,MM94,DOT96,DM97},
it may be that the true phase diagram is more complicated than anticipated, with the occurrence of an intermediate ``glassy'' phase
signaled by a breaking of symmetry in the replica formalism. The fact that the RFIM, for a certain range of temperatures and fields, has 
a complicated energy landscape (so that the mean-field equations for the local magnetizations have
many solutions \cite{LMP95}) also explains the failure of standard renormalization group perturbation theory \cite{Y77,PS81}
 and the breakdown of dimensional reduction. 

In this work, we propose a theory for the RFIM that allows an approximate study of the influence of dimensionality
on equilibrium properties. This is done by extending to this model the so-called self-consistent 
Ornstein-Zernike approximation (SCOZA) developed by Hoye and Stell \cite{HS77} for simple fluid and lattice-gas
systems. Recently, the SCOZA has been shown to provide a very good description of the properties
of the three-dimensional Ising model, even in the close vicinity of the critical point \cite{DS96}. We expect that the same approximation 
scheme will be useful in the case of disordered systems, as  well. In a previous work  \cite{KRT97}, we 
studied as a first application the site-diluted Ising model, showing that this approach can indeed provide an accurate description of the dependence of 
the critical temperature on dilution. To overcome the lack of translational invariance of the Hamiltonian due to the random disorder, 
we applied the replica method in an unusual way, replacing the original system by $n+1$ coupled systems with translationally invariant 
interactions. In the following, we shall use the same procedure and assume that there is no violation of replica symmetry. Our formalism can
be generalized to study a possible replica symmetry breaking, but this question will be adressed  in a forthcoming paper dealing also 
with the spin glass problem \cite{KRT98}.

The paper is organized as follows. In section II, we derive the replica-symmetric Ornstein-Zernike equations for the RFIM.  In section III, we 
introduce successively  the Random Phase approximation (which is just another way of obtaining the mean-field thermodynamics) and the
Optimized Random Phase approximation, which represents a first improvement on mean-field theory for hard-spin systems in finite 
dimensions. We then derive in section IV
the SCOZA partial differential equations for the Gaussian and bimodal 
distributions of the random field. In section IV, we analyze the solution in terms of high-temperature and $1/d$ series expansions.
A summary and a discussion are provided in section V.

\section{Replica-symmetric Ornstein-Zernike equations}

The RFIM is defined by the Hamiltonian

\be
{\cal H} = -J \sum_{<ij>} \sigma_{i}\sigma_{j} -\sum_i (H+h_i )\sigma_i 
\ee
where  $J >0$, $\sigma_i=\pm 1$, and $<ij>$ indicates that the sum is over pairs of nearest neighbor sites of a d-dimensional lattice. $H$ is a uniform
magnetic field and the local fields $h_i$ are independent random variables distributed according to some common probability 
distribution ${\cal P}(h)$. 
Quenched thermodynamic averages are defined by

\be
\overline{<A>_T}=\overline{Tr [\exp(-\beta {\cal H})A]/Tr [\exp(-\beta {\cal H})]}
\ee
where $\beta=1/(k_BT)$ and the overbar denotes an average over the random field distribution. If not stated otherwise, 
we shall concentrate on the class of distributions that depend on a single positive parameter $h_0$ which measures the strength of disorder. 
They  thus satisfy

\be
{\cal P}(h) dh={\cal P}(\tau) d\tau
\ee
where $\tau=h/h_0$ and ${\cal P}(\tau)$  is independent of $h_0$. This includes the much studied Gaussian distribution, 

\be
{\cal P}(h)=(2\pi h_0^2)^{-1/2} \exp[-h^2/(2h_0^2)] \ ,
\ee
and the bimodal distribution,

\be
{\cal P}(h)=\frac{1}{2}[\delta(h-h_0)+\delta(h+h_0)] \ .
\ee
In the first case, the quenched variables $\tau_i$ are continuous (soft) spin variables whereas they only take the values $\pm 1$ (Ising-like)
in the second case. The crucial point is that both quenched and annealed variables, $\tau_i$ and $\sigma_i$ respectively, are present. 
Such quenched-annealed two-species systems 
have been previously introduced in the context of liquid-state theory to describe continuum fluids adsorbed in porous media  
\cite{MG88,G92}.
The replica method was used to obtain a set of exact equations for the pair correlation functions \cite{GS92}
and to derive thermodynamic relations \cite{RTS94}. The same technique can be used here.
We first introduce $n$ copies of the annealed spin variables and consider the Hamiltonian 

\be
{\cal H}_n= -J \sum_{<ij>,a}\sigma_{i}^a\sigma_{j}^a -h_0\sum_{i,a}\tau_i \sigma_i ^a -
H\sum_{i,a}\sigma_i ^a \ .
\ee

Now, in contrast with what is usually done, we do {\it not} perform the average over the disorder variables
to get an effective Hamiltonian  (see however Appendix A). We treat the variables \{$\sigma_i^a$\} and \{$\tau_i$\} on an equal footing and we
consider ${\cal H}_{rep}(\{\sigma_i^a\},\{\tau_i\})={\cal H}_n-1/\beta \ \sum_i \ln {\cal P}(\tau_i)$
as the Hamiltonian of a mixture of $(n+1)$ spin species. The corresponding free energy is
${\cal F}_{rep}=-1/\beta \ \ln [Tr \exp(-\beta {\cal H}_{rep})]$,
where the trace is taken over \{$\sigma_i^a$\} and \{$\tau_i$\}. The average free energy of the RFIM is then given by

\be
{\cal F}=- \frac{1}{\beta} \lim_{n \rightarrow 0} \frac {1}{n}[\exp(-\beta {\cal F}_{rep})-1]  \ ,
\ee
and the average magnetization $m=\overline{<\sigma_i>_T}$ is given by
\be
m= \lim_{n \rightarrow 0} \frac {1}{n}\sum_a<\sigma_i^a>_{rep}
\ee
where $<...>_{rep}$ denotes the average with respect to ${\cal H}_{rep}$. The disconnected and 
connected correlation functions, $G_{dis}({\bf r})=\overline{<\sigma_{\bf 0}>_T<\sigma_{\bf r}>_T}-m^2$ and 
$G_{con}({\bf r})=\overline{<\sigma_{\bf 0}\sigma_{\bf r}>_T-<\sigma_{\bf 0}>_T<\sigma_{\bf r}>_T}$, are related to the replica correlation functions
$G^{ab}({\bf r})=<\sigma_{\bf 0}^a \sigma_{\bf r}^b>_{rep}-<\sigma_{\bf 0}^a>_{rep}<\sigma_{\bf r}^b>_{rep}$ ($a,b=1..n$) by

\be
G_{dis}({\bf r})=\lim_{n \rightarrow 0} \frac{1}{n(n-1)}\sum_{a\neq b}G^{ab}({\bf r})
\ee
and
\be
G_{con}({\bf r})=\lim_{n \rightarrow 0} \frac{1}{n}\sum_{a}G^{aa}({\bf r})-G_{dis}({\bf r}) \ .
\ee
We also introduce the correlation functions $G_{00}({\bf r})=\overline{\tau_{\bf 0}\tau_{\bf r}}-\overline{\tau}^2$ and 
$G_{01}({\bf r})=\overline{\tau_{\bf 0}<\sigma_{\bf r}>_T}-\overline{\tau} m$ which are related to the replica
correlation functions $G^{00}({\bf r})=<\tau_{\bf 0}\tau_{\bf r}>_{rep}-<\tau_{\bf 0}>_{rep}<\tau_{\bf r}>_{rep}$ and 
$G^{0a}({\bf r})=<\tau_{\bf 0}\sigma_{\bf r}^a>_{rep}-<\tau_{\bf 0}>_{rep}<\sigma_{\bf r}^a>_{rep}$ by

\be
G_{00}({\bf r})=\lim_{n \rightarrow 0} G^{00}({\bf r})
\ee
and
\be
G_{01}({\bf r})=\lim_{n \rightarrow 0} \frac{1}{n}\sum_{a}G^{0a}({\bf r}) \ .
\ee
One has $G_{00}({\bf r})=(\overline{\tau ^2}-\overline{\tau}^2)\ \delta_{{\bf r},{\bf 0}}$ for the distribution functions of 
uncorrelated random fields that are considered here (in principle, the case of correlated fields can be studied with the
present formalism as well). Because of the hard spin condition $\sigma_i=\pm 1$, one also has the sum-rule
 
\be
G_{11}({\bf r}={\bf 0})=1-m^2
\ee
where $G_{11}({\bf r})=G_{con}({\bf r})+G_{dis}({\bf r})$ (the Fourier transform of $G_{11}({\bf r})$ is the structure factor 
measured, e.g., in scattering experiments \cite{BY91}).
Eq. (13) is equivalent  to the so-called ``core'' condition in a lattice-gas \cite{S69}.
On the other hand,
\be
G_{con}({\bf r}={\bf 0})=1-q
\ee
where $q=\overline{<\sigma_i>_T^2}$ is the standard spin-glass order parameter.

We now provisionally assume that the external magnetic field is nonuniform, replica-dependent, and has an extra component 
$H_i^0$ that acts on the spin $\tau_i$. One has  $m_i^a=<\sigma_i^a>_{rep}=-\partial {\cal F}_{rep}/\partial H_i^a$ where 
$a$ now represents not only the $n$ replicas $1...n$ but also species $0$ and $\sigma_i^0$ and $m_i^0$ stand 
for $\tau_i$ and $<\tau_i>_{rep}$, respectively. The Legendre
 transform that takes the fields $H_i^a$ into $m_i^a$ defines the Gibbs potential 
\be
{\cal G}_{rep}={\cal F}_{rep}+\sum_{i}\sum_{a=0}^{n}H_i^a m_i^a
\ee
which satisfies
\be
H_i^a=\frac{\partial {\cal G}_{rep}}{\partial m_i^a}\ .
\ee
${\cal G}_{rep}$ generates the direct correlation functions (or proper vertices in field-theoretic language) which we define by
\be
C_{ij}^{ab}=\beta \frac{\partial ^2 {\cal G}_{rep}}{\partial m_i^a \partial m_j^b}  \ \ (a,b=0,1,...n) \ .
\ee
Since $G_{ij}^{ab}= -1/\beta \  \partial ^2 {\cal F}_{rep}/\partial H_i^a \partial H_j^b$, we have a set of Ornstein-Zernike equations
\be
\sum_{l} \sum_{c=0}^nG_{il}^{ac}C_{lj}^{cb}=\delta_{i,j}\delta_{a,b} \ ,
\ee
which, in the limit of a uniform replica-independent magnetic field, become in Fourier space

\be
\sum_{c=0}^n \hat{G}^{ac}({\bf k})\hat{C}^{cb}({\bf k})=\delta_{a,b} \ .
\ee
By taking the limit $n \rightarrow 0$ and assuming replica symmetry, we finally obtain

\begin{mathletters}
\begin{equation}
\hat{G}_{00}({\bf k})=\frac{1}{\hat{C}_{00}({\bf k})}
\end{equation}
                                   
\begin{equation}
\hat{G}_{con}({\bf k})=\frac{1}{\hat{C}_{con}({\bf k})}
\end{equation}

\begin{equation}
\hat{G}_{dis}({\bf k})=[\frac{\hat{C}_{01}^2 ({\bf k})}{\hat{C}_{00}({\bf k})}-\hat{C}_{dis}({\bf k})] \frac{1}{\hat{C}_{con}^2({\bf k})}
\end{equation}

\begin{equation}
\hat{G}_{01}({\bf k})=-\frac{\hat{C}_{01}({\bf k})}{\hat{C}_{00}({\bf k})\hat{C}_{con}({\bf k})} \ ,
\end{equation}
\end{mathletters}
where the $C$'s are related to the corresponding replica direct correlation functions by  relations similar to Eqs. (9-12). 
Note that the first equation decouples 
from the other ones, as it should be, and that $G_{01}=G_{10}$ and $C_{01}=C_{10}$ by symmetry.
These equations, hereafter called the replica-symmetric Ornstein-Zernike (RSOZ) equations, represent 
the starting point of our study. Apart from a few notational changes, they are the same as those
originally derived by Given and Stell \cite{GS92} for a quenched-annealed mixture in the context of liquid-state theory (see
also Refs. \cite{PRST95,PRT96} for an application to a lattice-gas model of a fluid in a disordered matrix). Because of the Legendre
transform, $m$ is now a control variable at our disposal instead of H, and in the following
all quantities will be considered as functions of the three independent variables $m, \tilde J=\beta cJ$ (where $c$ is the coordination 
number of the lattice) and $\tilde h_0=\beta h_0$. We shall be especially concerned
with the behavior of the susceptibility $\chi(m,\tilde J,\tilde h_0)=\partial m/ \partial (\beta H)$ given by 
\be
\chi(m,\tilde J,\tilde h_0)=\hat{G}_{con}({\bf k}={\bf 0})=\frac{1}{\hat{C}_{con}({\bf k}={\bf 0})} \ .
\ee
In the approximate theories that are discussed below, the divergence of $\chi$ at fixed $h _0$ defines the spinodal line in the $T-m$ plane. 
For a symmetric distribution (i.e., ${\cal P}(h)={\cal P}(-h)$),
and if the transition is continuous, the critical point is reached when the spinodal meets the magnetization curve at $m=0$. Alternatively, we 
can locate the critical temperature by plotting $\chi ^{-1}(m=0)$ as a function of $\tilde J$ at fixed $h _0$ .

\section{Random Phase and Optimized Random Phase approximations}

\subsection{Random Phase Approximation}

When one turns off the exchange interaction in the RFIM Hamiltonian $\cal{H}$, all quenched-averaged quantities
can be calculated straightforwardly by direct averaging over the random field distribution. 
Hereafter, we shall call the system where $J=0$ the reference system. One then has
\be
m=\overline{\tanh \beta (h_0 \tau +H)}
\ee
which can be inverted to get $ \beta H$ as a function of $m$ and $\tilde h_0$ (at least under the form of an infinite series).
 Then
\be
q_{ref}=\overline{\tanh ^2\beta (h_0 \tau +H)}
\ee
becomes a function of $m$ and $\tilde h_0$. For instance,
\be
q_{ref}(m,\tilde h_0)=m^2+(1-m^2)^2[\tilde h_0^2-2(1-2m^2)\tilde h_0^4+\frac{1}{3}(81m^4-90m^2+17)\tilde h_0^6+O(\tilde h_0^8)]
\ee
for the Gaussian distribution, and
\be
q_{ref}(m,\tilde h_0)= m^2+(1-m^2)^2[\tilde h_0^2-\frac{2}{3}(3m^2+1)\tilde h_0^4+\frac{1}{45}(225 m^4+30m^2+17)\tilde h_0^6
+O(\tilde h_0^8)]
\ee
for the bimodal distribution.

In the reference system, all correlation functions are purely local,
\be
G_{con}^{ref}({\bf r})=(1-q_{ref}) \ \delta_{{\bf r},{\bf 0}}
\ee
\be
G_{dis}^{ref}({\bf r})=(q_{ref}-m^2) \ \delta_{{\bf r},{\bf 0}}
\ee
\be
G_{01}^{ref}({\bf r})=v_{ref}\ \delta_{{\bf r},{\bf 0}} 
\ee
where $v_{ref}(m,\tilde h_0)=\overline{\tau \tanh \beta(h_0 \tau+H)}- \overline{\tau}m$. From the RSOZ equations, we 
then get

\be
C_{con}^{ref}({\bf r})=\frac{1}{1-q_{ref}} \ \delta_{{\bf r},{\bf 0}}
\ee
\be
C_{dis}^{ref}({\bf r})=\frac{v_{ref}^2c_{00}-q_{ref}+m^2}{(1-q_{ref})^2} \ \delta_{{\bf r},{\bf 0}}
\ee
\be
C_{01}^{ref}({\bf r})=- \frac{v_{ref} \ c_{00}}{1-q_{ref}}\ \delta_{{\bf r},{\bf 0}}
\ee
where $c_{00}=(\overline{\tau ^2}-\overline{\tau}^2)^{-1}$ ($c_{00}=1$ for the symmetric Gaussian and bimodal distributions defined
by Eqs. (4) and (5)).
Note that a special feature of the bimodal distribution is that $C_{dis}^{ref}({\bf r})=0 $, as can be readily checked.

Once a reference system has been chosen, the Random Phase approximation (RPA) 
consists in adding to the direct correlation fonctions of the reference system the pair interactions which had been turned off  
(see, e.g.,  Ref. \cite{HMcD76}).
In the present case, since there is no direct interaction between distinct replicas in the Hamiltonian ${\cal H}_{rep}$ (because
the average over disorder has not been performed explicitly), the RPA in Fourier space writes

\begin{mathletters}
\be
\hat{C}_{con}^{RPA}({\bf k})=\hat{C}_{con}^{ref}({\bf k})- \tilde J\ \hat{\lambda}({\bf k})
\ee
\be
\hat{C}_{dis}^{RPA}({\bf k})=\hat{C}_{dis}^{ref}({\bf k})
\ee
\be
\hat{C}_{01}^{RPA}({\bf k})=\hat{C}_{01}^{ref}({\bf k})
\ee
\end{mathletters}
where $\hat{\lambda}({\bf k})=1/c \sum_{{\bf e}}\exp(i {\bf k}.{\bf e})$ is the characteristic function of the lattice and ${\bf e}$
denotes a vector from the origin to one of its nearest neighbors. Using Eqs. (21) and (29), this leads to
\be
\chi_{RPA}=\frac{1-q_{ref}}{1-z_{RPA}}
\ee
where $z_{RPA}=\tilde J (1-q_{ref})$. The same result is obtained by differentiating the mean-field expression of
the magnetization \cite{SP77},
\be
m=\overline{\tanh \beta(cJ m+h_0\tau+H)} \ ,
\ee
with respect to $\beta H$. 
Therefore, the above RPA description is equivalent to the standard mean-field theory of the RFIM \cite{SP77,A78}. Note however that the RPA free energy
 is identical to the mean-field free energy only when it is obtained by integration of the susceptibility. 
As will be stressed in section IV, there are several routes to obtain the thermodynamics from the correlation functions, and when the latter
 are only known approximately
the different routes may not lead to the same results. For spin systems, the free energy or the Gibbs potential can be computed either
from the susceptibility given by Eq. (21) or by integrating with respect to temperature the enthalpy which is itself expressed in terms of the pair 
correlation functions. For simplicity, we only discuss in this section the results of the susceptibility route.

Then, for a symmetric distribution, and if  the transition is second-order, the boundary between the ferromagnetic and paramagnetic phases 
is given in the RPA by 
$z_{RPA}(m=0,\tilde J_c, \tilde h_0)=1$, i.e., 
\be
\tilde J_c [1-q_{ref}(m=0,\tilde h_0)]=1 \ .
\ee
In this approximation, the transition remains second-order along the whole phase boundary for the Gaussian distribution \cite{SP77}. 
On the other hand, it becomes first-order at sufficiently large disorder strength 
when the random field distribution has a relative minimum at zero field \cite{A78}. For instance, the tricritical point occurs at 
$\tilde J_t=3/2, \tanh^2(\tilde h_{0,t})=1/3$ for the bimodal distribution. Whether
this scenario still holds in finite dimensions (i.e. when $c$ is small) is an open question.

\subsection{Optimized Random Phase Approximation}

One shortcoming of the RPA is that the sum-rule, Eq. (13), is not satisfied. Indeed, from the RSOZ equations, we have

\be
\hat{G}_{11}^{RPA}({\bf k})=\frac{1-q_{ref}}{1-z_{RPA}\ \hat{\lambda}({\bf k})}+\frac{q_{ref}-m^2}{[1- z_{RPA}\ \hat{\lambda}({\bf k})]^2} 
\ee
(this expression leads to the well-known Lorentzian and Lorentzian-squared terms in the mean field structure factor near $k=0$ \cite{BY91}).
Introducing  the lattice Green's function \cite{J72}

\be
P({\bf r},z)=\frac{1}{(2\pi)^d}\int_{-\pi}^{\pi}d^d{\bf k}\frac{e^{i{\bf k}.{\bf r}}}{1-z\hat{\lambda}({\bf k})}
\ee
and going back to real space, we get

\be
G_{11}^{RPA}({\bf r})=(1-q_{ref})P({\bf r},z_{RPA})+(q_{ref}-m^2)[P({\bf r},z_{RPA})+z_{RPA}P'({\bf r},z_{RPA})]
\ee
where $P'({\bf r},z)\equiv \partial P({\bf r},z)/\partial z$. We thus have
\be
G_{11}^{RPA}({\bf r}={\bf 0})=(1-q_{ref})P(z_{RPA})+(q_{ref}-m^2) [P(z_{RPA})+z_{RPA}P'(z_{RPA})]
\ee
where $P(z)\equiv P({\bf r}={\bf 0},z)$. In general, this is different from $1-m^2$.

To cure this problem, one may simply add to $C_{con}^{ref}({\bf r})$ a state- and field-dependent perturbation 
potential that is different from zero 
at $ {\bf r}={\bf 0}$
and chosen in such a way that Eq. (13) is satisfied (of course, in the true system, the observables cannot depend on 
the value of the spin-spin interaction at  ${\bf r}={\bf 0})$. Using the terminology of  liquid-state theory \cite{HMcD76}, we call this 
approximation the Optimized Random Phase approximation (ORPA) \cite{AC72}. We thus write
\be
\hat{C}_{con}^{ORPA}({\bf k})=c_c(m,\tilde J,\tilde h_0)- \tilde J \ \hat{\lambda}({\bf k})
\ee
whereas Eqs. (32b) and (32c) are left unchanged. Introducing  $z_{ORPA}=\tilde J/c_c$, we now have
\be
\hat{G}_{11}^{ORPA}({\bf k})=\frac{1/c_c}{1-z_{ORPA}\ \hat{\lambda}({\bf k})}+\frac{q_{ref}-m^2}{(1-q_{ref})^2}\ \frac{1/c_c^2}
{[1- z_{RPA}\ \hat{\lambda}({\bf k})]^2}
\ee
and the sum-rule, Eq. (13), writes
\be
\frac{z_{ORPA}}{\tilde J} P(z_{ORPA}) +\frac{q_{ref}-m^2}{(1-q_{ref})^2} (\frac{z_{ORPA}}{\tilde J})^2
[P(z_{ORPA})+z_{ORPA}P'(z_{ORPA})]=1-m^2
\ee
which is viewed as an implicit equation for $z_{ORPA}(m,\tilde J,\tilde h_0)$. 

The susceptibility $\chi_{ORPA}=z_{ORPA}/{[\tilde J}(1-z_{ORPA})]$
diverges when $z_{ORPA}=1$. For a symmetric distribution, and if the transition is second-order, the inverse 
critical temperature is then given by

\be
\frac{P(1)}{\tilde J_c}+\frac{q_{ref}(m=0,\tilde h_0)}{[1-q_{ref}(m=0,\tilde h_0)]^2}\frac{P(1)+P'(1)}{\tilde J_c^2}=1 \ .
\ee 
In the absence of quenched disorder (i.e. when $h_0=0$), one has $q_{ref}=m^2$ and the inverse critical  temperature is given by 
$\tilde J_c= P(1)$, i.e., $\beta_c= P(1)/(cJ)$. In
this case, the ORPA is identical to the mean-spherical approximation (MSA) \cite{LP66}. (The MSA can be extended to the RFIM 
by choosing as reference the system where $J=0$ {\it and } $h _0=0$; then, Eqs. (32b) and (32c) are replaced by 
$\hat{C}_{dis}^{MSA}({\bf k})=0$ and $\hat{C}_{01}^{MSA}({\bf k})=-\tilde h_0$, which amounts to a linearization of the ORPA expressions
with respect to $\tilde h_0$.)

For $d \rightarrow \infty$, $P(z)  \rightarrow 1$ and  $P'(z)  \rightarrow 0$, and we recover from Eq. (43) the mean-field equation 
for the critical temperature, Eq. (35). On the other hand, for $d \le 4$, $P'(z)$ diverges at $z=1$ 
(whereas $P(1)$ diverges for $d \le 2$), and the critical temperature, solution of Eq. (43), is driven to zero. This unfortunate drawback is shared by any 
Ornstein-Zernike theory that assumes that $C_{con} ({\bf r})$ has the same range as the exchange interaction and that makes use of the
susceptibility route. In this case, one finds from the 
RSOZ equations that the critical exponents $\eta$ and $\bar{\eta}$, defined by $G_{con}({\bf r}) \sim r^{-d+2-\eta}$ and $G_{dis}({\bf r}) 
\sim r^{-d+4-\bar{\eta}}$  when $r \rightarrow \infty$ at the critical point, are both zero. As a consequence, 
there is no critical point at nonzero temperature for $d \le 4$ because this would lead to the unacceptable result that  
$G_{11}({\bf r})$ does not decrease to zero at long distances.

By using the expansion of $P(z)$ about  $z =1$ for $d>2$ and $d \neq 4,6,8...$ \cite{J72}, 

\be
P(z)=[P(1)+b_1(1-z)+b_2(1-z)^2+...]+(1-z)^{d/2-1}[c_0+c_1(1-z)+c_2(1-z)^2+...]
\ee
(for $d =4,6,8,...$, there are logarithmic corrections), it is easily found from Eq. (42) that for $4 < d < 6$ the ORPA gives rise to the critical 
exponents of the random-field spherical model \cite{LT74}, i. e. $\nu=1/(d-4)$,
 $\gamma=2/(d-4)$ and $\delta=d/(d-4)$. The mean-field exponents are recovered for $d \geq 6$.
Therefore the ORPA susceptibility route leads to a $d \rightarrow d-2$ dimensional reduction that is a direct consequence of 
the Ornstein-Zernike approximation for the direct correlation functions. On the other hand, it is easy to show that the enthalpy route
yields classical (mean-field) critical exponents. Both routes give back the mean-field results in the limit $d \rightarrow \infty$
(a more complete study of the ORPA for hard-spin systems with quenched  disorder will be presented elsewhere \cite{KRT98b}). 

An  illustration of the predictions of the ORPA  is given in Tables I  and II for the 5-d hypercubic lattice ($c=10$, $P(1)=1.156308$,
$P'(1)=0.778633$). The  inverse critical temperatures obtained from Eq. (43) are compared to the best available estimates obtained from a
fifteen-term high-temperature series expansion of the susceptibility \cite{GAAHS93} (the disorder strength is measured here 
in terms of the reduced variable $\tilde g=h_0^2/(cJ)^2=\tilde h_0^2/\tilde J^2$; this differs from the $g$ defined in Ref. \cite{GAAHS93} by
the factor $1/c^2$). Although the ORPA represents a clear improvement on mean-field theory, we see that the agreement with the ``exact'' 
results deteriorates significantly as the field
increases. For the bimodal distribution, Eq. (43) has no solution when $\tilde g \geq 0.1199$ and one may suspect that 
the transition becomes first-order for the highest values of the field. In order to decide on
the existence of a tricritical point, we expand the inverse susceptibility around $m=0$ along the critical isotherm and look for the change
of sign of $\chi^{-1}/\mid m\mid ^{\delta -1}$ at $m=0$ and $\tilde J=\tilde J_c$.
Since $\chi^{-1}_{ORPA} \sim (1-z_{ORPA})$ when $z_{ORPA} \rightarrow 1^-$, this amounts to expanding the solution of Eq. (42). Using 
the expansion given in Eq. (44), we find after some calculations that the two conditions for tricriticality are Eq. (43) and

\be
\frac{A_2(\tilde h_0)}{\tilde J_c^2}[P(1)+P'(1)] =-1
\ee
where $A_2(\tilde h_0)$ is the coefficient of $m^2$ in the expansion of $(q_{ref}-m^2)/(1-q_{ref})^2$ around $m=0$, 

\be
A_2(\tilde h_0)=\frac{(1+\overline{t^2})\overline{(1-t^2)(1-3t^2)}}{(1-\overline{t^2})^5} -\frac{1}{(1-\overline{t^2})^2}
\ee
where $t=\tanh (\tau \tilde h_0)$. For the  Gaussian distribution, Eqs. (43) and (45) have no solution and no tricritical point
appears in finite dimension. On the other hand, there is always a solution for the bimodal distribution, and one finds for instance 
 $\tilde g_t^{ORPA}=0.117$ and 
$\tilde J_t^{ORPA}=1.85$ for the 5-d hypercubic lattice. By comparing  to the mean-field predictions, $\tilde g_t^{RPA}=0.193$ 
and $\tilde J_t^{ORPA}=1.5$, we see that there is a range of fields where the transition is driven first order by fluctuations, 
in agreement with the conclusion of a previous  high-temperature series analysis  \cite{H85}. 
For more general symmetric random-field distributions, 
a careful analysis of Eqs. (43) and (45) shows that  the conclusion of Aharony \cite{A78} based on
mean-field theory remains unchanged for the ORPA: a sufficient condition for the occurence of a tricritical point is that ${\cal P}(h)$
has a minimum at zero field.

\section{Self-Consistent Ornstein-Zernike approximation}

As is well known in liquid-state theory and has been mentionned above, solving the 
Ornstein-Zernike equations with an approximate expression of the direct correlation 
functions like in the RPA, the ORPA  or any other approximate closure relation such as  the Percus-Yevick approximation or 
the hypernetted chain equation \cite{HMcD76},
generally leads to thermodynamic inconsistencies. In the language of magnetic systems, this means that different Gibbs potentials are 
obtained depending on whether one uses the susceptibility or the enthalpy routes (the enthalpy is defined by  ${\cal E}={\cal G}+T{\cal S}={\cal U}
+MH$ where ${\cal S}$ and ${\cal U}$ are the entropy and internal energy, respectively, and $M=Nm$ is the total magnetization).
For instance, in the pure Ising model, the former route corresponds to the double integration of 
the equation $\chi^{-1}=\hat C({\bf k}={\bf 0})=\partial ^2 (\beta{\cal G}/N)/\partial m^2$ with respect to $m$ (at constant $T$), and the
later to the integration of the Gibbs-Duhem relation ${\cal E}=\partial \beta {\cal G}/\partial \beta $ with respect to $\beta$ (at constant $m$).
The requirement that the two routes lead to the same results is thus embodied in the relation
$\partial \hat C({\bf k}={\bf 0})/\partial \beta =\partial ^2({\cal E}/N)/\partial m^2$, provided that the appropriate initial conditions are 
satisfied. Since the enthalpy per spin is given by
 ${\cal E}/N=-J/N \sum_{<ij>}<\sigma_i \sigma_j>_T=-cJ/2 [G({\bf r}={\bf e})+m^2]$, this relation, together with the 
Ornstein-Zernike equation, may be viewed as an exact equation for the pair correlation function. 

The extension to the RFIM is straightforward. Let us consider the general variation of the average free energy due to variations of the three
control parameters $J, h_0$ and $H$. We have
\be
d{\cal F}=-dJ \sum_{<ij>}\overline{<\sigma_i \sigma_j>_T} -dh_0\sum_i\overline{\tau_i<\sigma_i>_T}-dH\sum_i\overline{<\sigma_i>_T} \ .
\ee
Then, from the definitions of $G _{11}({\bf r})$ and $G_{01}({\bf r})$, we readily find that
\be
d(\beta{\cal G}/N)=-\frac{1}{2}(G_{11}({\bf r}={\bf e})+m^2)d\tilde J -(G_{01}({\bf r}={\bf 0})+{\bar \tau} m)d \tilde h_0+\beta Hdm \ .
\ee
On the other hand, we have from Eq. (17)
\be
\hat C_{con}({\bf k}={\bf 0})=\frac{\partial ^2 (\beta{\cal G}/N)}{\partial m^2} \ ,
\ee
so that we get  the three ``Maxwell relations'':
\be
\frac{\partial \hat C_{con}({\bf k}={\bf 0})}{\partial \tilde J}=-1 -\frac{1}{2}\frac{\partial ^2 G_{11}({\bf r}={\bf e})}{\partial m^2} \ ,
\ee
\be
\frac{\partial G_{01}({\bf r}={\bf 0})}{\partial \tilde J}=\frac{1}{2}\frac{\partial G_{11}({\bf r}={\bf e})}{\partial \tilde h_0} 
\ee
\be
\frac{\partial \hat C_{con}({\bf k}={\bf 0})}{\partial \tilde h_0}=-\frac{\partial ^2G_{01}({\bf r}={\bf 0})}{\partial m^2} \ .
\ee
Since the enthalpy density is given from the Hamiltonian $\cal{H}$ by
\be
{\cal E}/N=-\frac{cJ}{2}[G_{11}({\bf r}={\bf e})+m^2]-h_0[G_{01}({\bf r}={\bf 0})+{\bar \tau} m] \ ,
\ee
the combination of Eqs. (50) and (52) yields
\be
\frac{\partial \hat C_{con}({\bf k}={\bf 0})}{\partial \beta} =\frac{\partial ^2({\cal E}/N)}{\partial m^2} \ ,
\ee
which generalizes the self-consistency relation for the pure Ising model discussed above and used in Ref. \cite{DS96}.

The SCOZA strategy is now the following. We assume that the three direct correlation functions $C_{con}({\bf r}),C_{dis}({\bf r})$
and $C_{01}({\bf r})$ have the same range as in the RPA but with values (for the present problem at ${\bf r}={\bf 0}$  and/or ${\bf r}={\bf e}$)
which are state- and field-dependent. We thus write in Fourier space

\begin{mathletters}
\be
\hat C_{con}^{SCOZA}({\bf k})=c_c(m,\tilde J,\tilde h_0)[1-z(m,\tilde J,\tilde h_0)\hat \lambda({\bf k})]
\ee
\be
\hat C_{dis}^{SCOZA}({\bf k})=c_d(m,\tilde J,\tilde h_0)
\ee
\be
\hat C_{01}^{SCOZA}({\bf k})=c_{01}(m,\tilde J,\tilde h_0)
\ee
\end{mathletters}
(for notational simplicity, we do not use the index SCOZA for $z,c_c,c_d$ and $c_{01}$). It follows from the RSOZ equations that

\begin{mathletters}
\be
G_{con}^{SCOZA}({\bf r})=\frac{1}{c_c}P({\bf r},z)
\ee
\be
G_{dis}^{SCOZA}({\bf r})=\frac{c^2_{01}/c_{00}-c_d}{c_c^2}\frac{\partial}{\partial z}[zP({\bf r},z)]
\ee
\be
G_{01}^{SCOZA}({\bf r})=-\frac{c_{01}}{c_c \ c_{00}}P({\bf r},z) \ .
\ee
\end{mathletters}
We then impose that the exact relations, Eq. (13) and Eqs. (50-52), be satisfied.
This leads to a set of partial differential equations (PDE)
in the unknown quantities $z,c_c,c_d$ and $c_{01}$. This is not enough, however, because only two of Eqs. (50-52) are independent provided
that the appropriate initial condifions are satisfied
(for instance Eqs. (50) and (51)). We thus need an additional relationship between the pair correlation functions.

This extra equation is readily found in  the case of the Gaussian probability distribution which has the special property that
$\int A\ \tau {\cal P}(\tau) d\tau=\int (dA/d\tau) {\cal P}(\tau) d\tau$. Choosing $A=<\sigma_i>_T$ yields
$\overline{<\sigma_i>_T\tau_j}=\overline{d<\sigma_i>_T/d\tau_j}=\beta h_0\ \overline{<\sigma_i\sigma_j>_T-<\sigma_i>_T<\sigma_j>_T}$, and
one gets the exact relationship
\be
G_{01}({\bf r})=\tilde h_0 \ G_{con}({\bf r}) \ .
\ee
Since $c_{00}=1$, this gives, when inserted in the RSOZ equations,
\be
\hat C_{01}({\bf k})=-\tilde h_0 \ . 
\ee
We thus take $c_{01}=-\tilde h_0 $ in the SCOZA equation (55c) and we use the sum-rule, Eq. (13), to express $c_d$ as a function of $c_c$,

\be
c_d=\tilde h_0^2+c_c\frac{P(z)-c_c(1-m^2)}{P(z)+zP'(z)}\ .
\ee
Using the fact that the Green's function at nearest-neighbor separation satisfies $P({\bf r}={\bf e},z)=[P(z)-1]/z$, we 
finally  obtain from Eqs. (50) and (51) two coupled PDE's in $z(m,\tilde J,\tilde h_0)$ and $c_c(m,\tilde J,\tilde h_0)$,

\begin{mathletters}
\be
\frac{\partial}{\partial \tilde J} [c_c(1-z)]=-1-\frac{1}{2}\frac{\partial ^2}{\partial m^2}
[ \frac{1}{c_c}\frac{P(z)-1}{z}+(1-m^2- \frac{P(z)}{c_c})\frac{P'(z)}{P(z)+zP'(z)}]
\ee
\be
\frac{\partial}{\partial \tilde J}[\frac{P(z)}{c_c}]=\frac{\partial}{\partial (\tilde h_0^2)}
[\frac{1}{c_c}\frac{P(z)-1}{z}+(1-m^2- \frac{P(z)}{c_c})\frac{P'(z)}{P(z)+zP'(z)}] \ .
\ee
\end{mathletters} 
When setting $c_c=P(z)/(1-m^2)$ in these two equations, the first one reduces to the PDE for the pure nearest-neighbor lattice-gas ($h_0=0$,
with the usual substitution $\rho=(1+m)/2$ and $w=4J$)
which has been studied in Ref. \cite{DS96}; see also the discussion in section V. In fact, the exact relationship, Eq. (57),
 suggests that  it was unnecessary to introduce the correlation function 
$G_{01}({\bf r})$ in the Gaussian case and that one could have averaged over the quenched disorder from the outset, as
is usually done. We show in Appendix A that the same expressions of the correlation functions
and of the thermodynamic quantities are obtained within the SCOZA when one uses this alternative route.

The case of the bimodal distribution is somewhat more complicated. As already noticed, it also has a special property, namely,
 $C_{dis}^{ref}({\bf r})=0$. However, $C_{dis}({\bf r})$ is not zero when $J \neq 0$ and  there is no reason {\it a priori}
to set $c_d(m,\tilde J,\tilde h_0)=0$  in Eq. (55b). The solution to this problem consists in introducing an additional independent variable 
that allows to control the mean value of the random-field. Indeed, Eq. (17) tells us that
\be
C_{ij}^{0a}=\beta\frac{\partial ^2 {\cal G}_{rep}}{\partial m_i^0 \partial m_j^a} 
\ee
which gives, in the limit $n \rightarrow 0$ and for a uniform magnetic field, 
\be
\hat C_{01}({\bf k}={\bf 0})=\frac{\partial ^2 (\beta {\cal G}/N)}{\partial \bar{\tau}\ \partial m} \ .
\ee
In consequence, we derive from Eq. (48) two additional Maxwell relations
 
\be
\frac{\partial \hat C_{01}({\bf k}={\bf 0})}{\partial \tilde J}=-\frac{1}{2}\frac{\partial ^2 G_{11}({\bf r}={\bf e})}{\partial \bar{\tau} \ \partial m}
\ee
\be
\frac{\partial \hat C_{01}({\bf k}={\bf 0})}{\partial \tilde h_0}=-1-\frac{\partial ^2G_{01}({\bf r}={\bf 0})}{\partial \bar{\tau}\ \partial m}\ .
\ee
Only three of Eqs. (50-52) and Eqs. (63-64) are independent but we now have with Eq. (13) the right number of equations to
determine unambiguously $z,c_c,c_d$, and $c_{01}$. In Appendix B, we apply this procedure to the asymmetric bimodal probability distribution
 ${\cal P}(h)=p\delta(h-h_0)+(1-p)\delta(h+h_0)$. Since $\bar{\tau}= 2p-1$, the parameter $p$ can be varied independently of  
$m,\tilde J$ and $\tilde h_0$ to change $\bar{\tau}$. This leads to three PDE's, Eqs. (B3). 
Of course, it is significantly more difficult to solve three PDE's than only two and it is highly desirable
to simplify the problem, especially if one is only interested in the case $p=1/2$. Fortunately, it turns out that $c_d=0$ is a very good 
approximation to the full solution of Eqs. (B3) (more precisely, we show in Appendix B that $c_d=0$ through order $\beta^6$ 
in the high-temperature series expansion of the solution of Eqs. (B3)). This is probably related to the fact that one has 
 $C_{dis}^{ref}({\bf r})=0$ even when  $p\neq 1/2$.
With this approximation, we are left with only three unknown quantities, $z,c_c$ and $c_{01}$, which can be determined by using
Eq. (13) and Eqs. (50-51). Returning to the case $p=1/2$ and introducing $f(m,\tilde J,\tilde h_0) =-c_{01}/c_c$ as a new unknown 
variable, we eliminate $c_c$ from Eq. (13) (cf. Eq. (B2) with $r=f$),

\be
c_c=\frac{P(z)}{1-m^2-f^2[P(z)+zP'(z)]} \ ,
\ee
and we finally obtain from Eqs. (50) and (51) two coupled PDE's in $z(m,\tilde J,\tilde h_0)$ and $f(m,\tilde J,\tilde h_0)$ 
(cf. Eqs. (B3a) and (B3b) with $r=f$),

\begin{mathletters}
\bea
\frac{\partial}{\partial \tilde J}[\frac{(1-z) P(z)}{1-m^2-f^2[P(z)+zP'(z)]}]=&-1&-\frac{1}{2}\frac{\partial ^2}{\partial m^2}
\{ [1-m^2-f^2(P(z)+zP'(z))]\frac{P(z)-1}{zP(z)}\nonumber\\
&+&f^2P'(z)\}
\eea
\be
\frac{\partial}{\partial \tilde J}[fP(z)]=\frac{1}{2}\frac{\partial}{\partial \tilde h_0}
\{ [1-m^2-f^2(P(z)+zP'(z))]\frac{P(z)-1}{zP(z)}+f^2P'(z)\}
\ee
\end{mathletters}

\section{Series expansions of the SCOZA solutions for the Gaussian and bimodal distributions}

To integrate the coupled PDE's, Eqs. (60) or (66), the initial values of $z, c_c$ or $f$ for $\tilde J=0$ must be known. 
The quantity $z$ is related to the second moment correlation length $\xi$ defined by $ \hat G_c({\bf k})= \hat G_c({\bf k}={\bf 0})[1-\xi^2k^2+...]$.
Specifically, one has $z=c/(c+\xi^{-2})$. Therefore $z=0$ for $\tilde J=0$  since all correlation functions are local in the reference system 
($\xi=0$). On the other hand,
$c_{c,ref}=1/(1-q_{ref})$ and in the bimodal case $f_{ref}$ is given by $f_{ref}^2=q_{ref}-m^2$. 
As discussed in section III.A, $q_{ref}$ can be obtained by direct averaging over the random-field distribution. 
It is much more convenient, however, to calculate the properties of the reference system
from the third ``Maxwell relation'', Eq. (52). When $J=0$, this equation leads to
\be
\frac{\partial \ c_{c,ref}(m,\tilde h_0)}{\partial \tilde h_0}=-\tilde h_0\frac{\partial ^2}{\partial m^2}\ \frac{1}{c_{c,ref}(m,\tilde h_0)}
\ee
for the Gaussian distribution, and 
\be
\frac{\partial}{\partial \tilde h_0}\frac{1}{1-m^2-f_{ref}^2(m,\tilde h_0)}=-\frac{\partial ^2 f_{ref}(m,\tilde h_0)}{\partial m^2} 
\ee
for the bimodal distribution. These equations are solved by imposing the boudary conditions at 
$m=0$ : $q_{ref}(0, \tilde h_0)=1/\sqrt(2\pi) \int_{-\infty}^{\infty} \exp(-u^2/2) \tanh^2(u\tilde h_0) du$ for the Gaussian distribution and 
$q_{ref}(0, \tilde h_0)=\tanh^2(\tilde h_0)$ for the bimodal distribution (morever, $f_{ref}$ and $c_{c,ref}$ 
are even functions of $m$). This is how we have obtained the series for $q_{ref}$ in Eqs. (24) and (25).

It may be noticed that Eqs. (60) and  (66) have a different behavior in the limit $h_0 \rightarrow 0$. Indeed,  $z$ and $c_c$ are even functions
of $h_0$ whereas $f$ is an odd function. Therefore, the solution of Eq. (66b) when $h_0 \rightarrow 0$ is $f=0$, which leads to 
$c_c=P(z)/(1-m^2)$ so that $z$ is identical to
the solution of the equation for the pure system that has been considered in  Ref. \cite{DS96}. On the contrary, in the Gaussian case, the 
solution of Eq. (60b) is not $c_c=P(z)/(1-m^2)$ when $h_0 \rightarrow 0$, and one does not recover the pure system results. 
The discrepancy, however, is extremely small.

In Ref. \cite{DS96}, the numerical integration of the SCOZA partial differential equation for the pure Ising model was performed by 
rewritting  this equation as a quasi-linear diffusion equation for which  implicit predictor-corrector
algorithms are available in the literature. The numerical  integration of the coupled PDE's, Eqs. (60) or (66), is more difficult and we 
defer this task to a later work. Note however that an interesting
feature of the theory is that a single run of integration steps sweeps the whole parameter space, so that the phase diagram in the 
$T-h_0$ plane can be obtained at once. In what follows, we only give some preliminary results obtained from series expansions. 

The SCOZA equations are indeed quite suitable for deriving high-temperature series. 
Since $z$ vanishes when $\tilde J \rightarrow 0$, one can expand  the Green's function $P(z)$ in powers of z,
\be
P(z)=1+\sum_{k\geq 2} P_k z^k
\ee
(with $P_2=1/c$), and substitute into the PDE's. We then express $z(m,\tilde J,\tilde h_0), c_c(m,\tilde J,\tilde h_0)$ and
$f(m,\tilde J,\tilde h_0)$ as triple series in $\tilde J, m$ and $\tilde h_0$,

\be
z(m,\tilde J,\tilde h_0)=(1-m^2)\sum_{i,j,k}z_{ijk}\tilde h_0^{2k} m^{2j}\tilde J^i
\ee

\be
c_c(m,\tilde J,\tilde h_0)=\sum_{i,j,k}c_{ijk}\tilde h_0^{2k}m^{2j}\tilde J^i
\ee

\be
f(m,\tilde J,\tilde h_0)=\sum_{i,j,k}f_{ijk}\tilde h_0^{2k+1} m^{2j}\tilde J^i \ ,
\ee
by using the fact that $z$ vanishes when $m=\pm 1$. Eventually, we  must gather in $\hat{C}_{con}({\bf k}={\bf 0})$ all terms at a 
given order in $\beta$ in order to obtain from Eqs. (21) and (55a) the high-temperature series expansion of the zero-field susceptibility. 
A careful analysis of the PDE's shows that to 
calculate $\chi(m=0)$ through the order $\beta^n$ it is sufficient to consider the triple series, Eqs. (70)-(72), up to finite values of $i,j,k$ :
$i_{max}=n$, $j_{max}=n-i$ and $k_{max}=n-i$ 
or $(n-i)/2$ for the Gaussian or bimodal distribution, respectively (this is for Eqs. (70) and (71); in Eq. (72), 
$k_{max}=n-i-1/2$ or $(n-i-1)/2$).
The crux of the calculation is that the coupled PDE's reduce at each order in $\tilde J,m,\tilde h_0$ to a system of linear  
algebraic equations in the unknown coefficients $z_{ijk}$ and $c_{c,ijk}$ or $z_{ijk}$ and $f_{ijk}$. In consequence,
the whole calculation can be performed with reasonable effort  using a symbolic computation software
like MAPLE or MATHEMATICA. The results for the hypercubic lattice in general dimension are

\bea
\chi(m=0) &=& 1+2d(\beta  J)+(-2d+4d^2-g)(\beta J)^2+ (4d/3-8d^2+8d^3-4dg)(\beta J)^3\nonumber\\
&+& (10d/3+16d^2/3-24d^3+16d^4+(4d-12d^2)g+2g^2)(\beta J)^4\nonumber\\
&+&(-28d/3+88d^2/5+24d^3-64d^4+32d^5+(-8d/3+24d^2-32d^3)g+10dg^2)(\beta J)^5 +...\nonumber\\
\eea
for the Gaussian distribution, and 
\bea
\chi(m=0) &=& 1+2d(\beta  J)+(-2d+4d^2-g)(\beta J)^2+ (4d/3-8d^2+8d^3-4dg)(\beta J)^3\nonumber\\
&+& (10d/3+16d^2/3-24d^3+16d^4+(4d-12d^2)g+2g^2/3)(\beta J)^4\nonumber\\
&+&(-28d/3+88d^2/5+24d^3-64d^4+32d^5+(-8d/3+24d^2-32d^3)g+14dg^2/3)(\beta J)^5 +...\nonumber\\
\eea
for the bimodal one (here $g=4d^2 \tilde g$ as in Ref. \cite{GAAHS93}). 
Comparison with the exact series \cite{GAAHS93} shows that the SCOZA series are exact through the fourth-order term
(at the order $\beta^5$, the only inexact coefficients are those of $d$ and $d^2$ which take the values -28/3 and 88/5 instead of -116/15 and 16,
 respectively).
More generally, the numerical values of the higher-order coefficients are in remarkable agreement with the exact ones. As an illustration, 
the terms of order $\beta^{15}$  are given in Appendix C  for both distributions (note that at each order in $\beta$ the coefficient of the highest-order term in $g$ is exact: this is because the reference system is treated exactly as a boundary condition to the PDE's). Similar series expansions can be
obtained for the structure factor $\hat G_{11}({\bf k}=0)$ and for other types of lattice (one only has to use
 the corresponding Green's function in the PDE's). 

We now turn to more quantitative predictions. Assuming that $\chi(m=0) \sim K\  (1-T/T_C)^{-\gamma}$ near $T=T_c$, we have performed a Dlog Pad\'e analysis of the high-T series
(since we hope to have a numerical solution of the PDE's in the near future, we have not tried to apply more sophisticated 
methods of analysis such as the ones used in Ref. \cite{GAAHS93}). 
For the critical temperature, the convergence between the different approximants is
fairly good and the results in $d=5$ using a $[5/6]$ approximant are given in Tables I and II (this approximant has been chosen 
because of its smooth behavior as a function of $\tilde g$). We see that the SCOZA predictions 
are in very good agreement with the estimates obtained from the exact high-temperature series expansion \cite{GAAHS93}. 
The phase diagrams displayed in Figures 1 and 2 show that the improvement on the RPA and the ORPA is indeed remarkable. 
On the other hand, the predictions for  $\gamma$ are rather sensitive to the order of the approximant and are therefore less reliable than for 
the critical temperature. $\gamma$ must anyhow be interpreted here as an effective exponent since we expect to obtain the random-field
spherical-model exponents asymptotically (this is not yet proved, however). The increase of $\gamma$ with $g$ has been noted in 
preceding studies \cite{GAAHS93} but for a serious discussion we must wait until the numerical solution of the PDE's is available.

To discuss the possible occurence of a tricritical point in hypercubic lattices, it is more convenient to consider $1/d$ expansions. These can be obtained either
by using  the $1/d$ expansion of the Green's function, $P(z)=1+z^2/(2d)+3z^4/(2d)^2+...$, or by using the conventional 
scaling $J=1/(2d)$ to reorganize the $\tilde J$-expansions. In the pure Ising model, this latter procedure readily yields the
$1/d$ expansion of the critical temperature around the mean-field $d \rightarrow \infty$ limit. In the RFIM, the calculation is 
more complicated because the
 mean-field critical temperature is itself solution of an implicit equation, Eq. (35). In consequence, we cannot use the conventional 
high-temperature
$\beta$-expansion as a starting point, but rather the $\tilde J$-expansion at constant $\tilde h_0=\beta h_0$. In order to get the systematic corrections around
mean-field theory, it is more convenient to take $q_{ref}(m=0)=\overline{\tanh ^2(\tilde h_0 \tau)}$ instead of $\tilde h_0$ 
as independent variable in the
PDE's. 
All this complicates the formal procedure, especially in the 
Gaussian case, and we have computed the expansion only through order 2. For the bimodal distribution, we find that the inverse zero-field
susceptibility is given by:
\bea
\chi^{-1}(m=0)=&&\frac{1}{1-x}-\tilde J+\frac{\tilde J^2}{2d}\ \frac{(1+x)(1-2x)}{1-x}+\frac{\tilde J^3}{(2d)^2}\frac{1}{1-x}[\frac{2}{3}(1+x)
(5x^2+2x-1)\nonumber\\
&&-\tilde J (18x^4-32x^3+12x^2+x-1)]+O(\frac{1}{d^3})
\eea
where $x\equiv q_{ref}(m=0,\tilde h_0)=\tanh^2(\tilde h_0)$. The coefficient
of the $1/d$ term can be shown to be exact.
The above expression yields  the expansion of $\tilde J_c$ around  the mean-field inverse critical temperature $\tilde J_c^{MF}$, solution of Eq. (35). 
The result is reproduced here only to first order:
\be
\frac{\tilde J_c}{\tilde J_c^{MF}}=1+\frac{1}{2d}\ \frac{1}{1-x_c^{MF}}\ \frac{(1+x_c^{MF})(1-2x_c^{MF})}{1-x_c^{MF}-2(\tilde gx_c^{MF})^{1/2}}
+O(\frac{1}{d^2})
\ee
where $x_c^{MF}=1-1/J_c^{MF}$. 
The tricritical point is obtained by solving simultaneously the equations $\chi^{-1}(m=0)=0$ and $\partial^2\chi^{-1}/\partial m^2\mid_{m=0}=0$ 
(since one always has classical mean-field critical exponents in a $1/d$ expansion, we take here $\delta=3$). This yields
\be
\tilde J_t=\frac{3}{2}+\frac{3}{2d}-\frac{23}{32d^2}+O(\frac{1}{d^3})
\ee
and
\be
\tilde g_t=0.1927-\frac{0.1320}{d}-\frac{0.4950}{d^2}+O(\frac{1}{d^3})
\ee
(recall that $\tilde g=\tilde h_0^2/\tilde J^2$). We thus approximately locate the tricritical point in $d=5$ at $\tilde J_t= 1.771$ and 
$\tilde g_t= 0.147$ (see Fig. 2). This confirms the conclusion reached with the ORPA: there is a range of field strengths
(for instance, $0.147< \tilde g < 0.193$ in 5-d) where the transition is driven first-order by fluctuations. The anomalous behavior of $\gamma$
observed at large $g$ in Table II is probably related to a crossover to this tricritical behavior. 

For the Gaussian distribution , we find
\bea
\chi^{-1}(m=0)=&&\frac{1}{1-x}-\tilde J-\frac{\tilde J^2}{2d}[1-x+\frac{xK(x)}{(1-x)^2}]
+\frac{\tilde J^3}{(2d)^2}\{\frac{2}{3}K(x)(1+x)\frac{K(x)-2(1-x)^2}{(1-x)^3}\nonumber\\
&&+\tilde J [\frac{x}{(1-x)^3}K^2(x)+\frac{x^2ü}{4(1-x)^2}\frac{dK^2(x)}{dx}+2xK(x)+(1-x)^3]\}+O(\frac{1}{d^3})
\eea
where $x=1/\sqrt(2\pi) \int_{-\infty}^{\infty} \exp(-u^2/2) \tanh^2(u\tilde h_0) du$ and $K(x)=(2\tilde h_0)^{-1}\ dx/d\tilde h_0$. The function 
$K(x)$ can be calculated once for all; one has $K(x)=1-4x+9x^2-24x^3+96x^4+O(x^6)$, and since $x =1-\sqrt(2/\pi)\tilde h_0^{-1}+O(\tilde h_0^{-2})$ when 
$\tilde h_0 \rightarrow \infty$ \cite{SP77}, $K(x) \sim (\pi/4)(1-x)^3$ when $x \rightarrow 1$. Here too, the expansion is exact to first order in $1/d$.
The resulting expansion of $\tilde J_c$ is:
\be
\frac{\tilde J_c}{\tilde J_c^{MF}}=1+\frac{1}{2d}[1+\frac{K(x_c^{MF})(x_c^{MF}+2 \tilde g)}{(1-x_c^{MF})^3-2\tilde g K(x_c^{MF})}]+O(\frac{1}{d^2})
\ee

The expansion given by Eq. (79) also permits to calculate the expansion of $g^*$, the variance of the random-field distribution for which $T_c=0$. We find
\be
g^*=\frac{2}{\pi}-\frac{1}{2d}(1+\frac{4}{\pi})-\frac{1}{(2d)^2}(\frac{1}{3}+\frac{6}{\pi})+O(\frac{1}{d^3}) \ .
\ee
Since there is no tricritical point in infinite dimension, we cannot perfom a perturbative expansion of $\tilde J_t$ or $\tilde g_t$
around mean-field theory. However, we can expand the equations $\chi^{-1}(m=0)=0$ and 
$\partial^2\chi^{-1}/\partial m^2\mid_{m=0}=0$ in powers of $1/d$ and look for a solution at a given value of $d$.
For $d\geq 5$ there is no indication that a tricritical point occurs. On the other hand, the calculation at order $1/d^2$ suggests that
the transition becomes first order for sufficiently strong disorder when $d\leq 4.2$. This is certainly a rough estimation which must be confirmed
by the numerical solution of the PDE's. The fact that $d \approx 4$ might be a ``critical'' dimension for the Gaussian RFIM has been also
suggested by preceding high-T series analysis \cite{H85}. 

Finally, we have also calculated the high-T and $1/d$ series expansion of the quantity $A$ defined as
\be
A=\lim_{T \rightarrow T_c} \frac{1}{\tilde h_0^2}\frac{\hat G_{dis}({\bf k}=0)}{\hat G_{con}^2({\bf k}=0)}=
\lim_{T \rightarrow T_c}\frac{1}{\tilde h_0^2}[\frac{\hat C_{01}^2({\bf k}=0)}{\hat C_{00}({\bf k}=0)}-\hat C_{dis}({\bf k}=0)] \ .
\ee
It has been argued in the literature \cite{SS85,MGN91} that $A$ should be equal to 1. In agreement with the analysis of Ref. \cite{GAAHS93},
we find that A is always finite (this is here a direct consequence of the OZ approximation for the correlation functions) and close, 
but not exactly equal to unity.

\section{Conclusion}

The calculations of the preceding section, based on the analysis of high-T and $1/d$ series expansions, shows that the 
thermodynamically self-consistent Ornstein-Zernike approximation (SCOZA) accurately predicts the critical temperature $T_c(h_0)$ of the RFIM in 
dimension $d > 4$ for both the Gaussian and bimodal distributions (at least in the range of disorder strength where we can
compare to the available ``exact'' results). 
For the bimodal distribution, we find that the phase transition becomes
first-order for sufficiently strong random fields, in agreement with mean-field theory \cite{A78}, and that the tricritical point moves to weaker fields 
as the dimension is reduced. A tricritical point may also appear in the Gaussian case for dimensions slightly above $4$, but the numerical solution
of the partial differential equations is needed to settle this question. This numerical solution, when available, will also permit to probe
the critical region and to calculate the effective exponents above the critical point (the analytical determination of the true asymptotic 
exponents, especially on the coexistence curve, and the elucidation of the scaling behavior of the theory are also challenging tasks). 
The accuracy of the theory shows that the assumption that the connected and disconnected direct correlation functions have the same range 
as in the RPA (i.e., essentially the same range as the interaction potentials) is quite reasonable, as long as one does not approach the critical point too closely or go to very low temperatures (the exact
structure of the correlation functions is certainly more complicated at low temperatures, as illustrated by the exact results in 1-d \cite{L92}).
The main shortcoming of this Ornstein-Zernike assumption is that $\eta=\bar \eta=0$ which forces $T_c$ to be zero for $d\leq 4$. 
On the other hand, the SCOZA can be easily generalized to the n-component version of the RFIM, as it has been done for the pure Ising
model \cite{HS97}. One can then show that the theory becomes exact in the spherical-model limit $n \rightarrow \infty$. This means
that the SCOZA should be even more accurate for the classical random-field X-Y or Heisenberg models. Finally, it must be stressed
that all the above results assume that replica symmetry is not broken in finite dimensions. This assumption must be justified
(or invalidated) by studying the stability of the replica symmetric solution. This will be done in a forthcoming paper devoted also to the 
application of the SCOZA to the Edwards-Anderson spin-glass model \cite{KRT98}.

\newpage

\appendix

\section{}

In the case of the Gaussian distribution, one usually averages $\exp(-\beta {\cal H}_n)$ over disorder to get an effective Hamiltonian

\be
{\cal H}_{eff}(\{\sigma_i^a\})= -J \sum_{<ij>,a} \sigma_{i}^a\sigma_{j}^a -\frac{\beta h_0^2}{2} \sum_{i,a,b}\sigma_i ^a\sigma_i ^b
-H\sum_{i,a}\sigma_i ^a \ .
\ee
The species $0$ does not appear any more but the replicas are now interacting. The corresponding RSOZ equations are

\begin{mathletters}
\begin{equation}
\hat{G}_{con}({\bf k})=\frac{1}{\hat{C}_{con}({\bf k})}
\end{equation}

\begin{equation}
\hat{G}_{dis}({\bf k})=-\frac{\hat{C}_{dis}({\bf k})}{\hat{C}_{con}^2({\bf k})} \ .
\end{equation}
\end{mathletters}

Working first in replica space and then taking the limit $n \rightarrow 0$, it is easily shown that the differential expression for the Gibbs potential
is now
\be
d(\beta{\cal G}/N)=-\frac{1}{2}(G_{11}({\bf r}={\bf e})+m^2)d\tilde J -\tilde h_0 G_{con}({\bf r}={\bf 0})d\tilde h_0+\beta Hdm
\ee
(${\bar \tau} = 0$ for the Gaussian distribution). Because of  Eq. (57), one thus gets the same Maxwell relations as Eqs. (50-52). 
On the other hand, some care must be taken in computing the internal energy or the enthalpy. The  internal energy 
 is obtained from ${\cal U}=\lim_{n \rightarrow 0} \ \partial {\cal U}_{eff}/\partial n$ where
 ${\cal U}_{eff}$ satisfies the Gibbs-Duhem relation ${\cal U}_{eff}=\partial(\beta {\cal F}_{eff})/\partial \beta$. 
Since the Hamiltonian ${\cal H}_{eff}$ is temperature-dependent, one  
has the unusual expression ${\cal U}_{eff}=<{\cal H}_{eff}+\beta \ \partial {\cal H}_{eff}/\partial \beta>_{eff}$ where $<..>_{eff}$ 
denotes the average with respect to  ${\cal H}_{eff}$. From Eq. (A1), this gives

\be
\frac{{\cal U}_{eff}}{N}=-\frac{cJ}{2} \sum_{a}[ G^{aa}({\bf r}={\bf e})+m_a^2] -\beta h_0^2 \sum_{a,b}[ G^{ab}({\bf r}={\bf 0})+m_a m_b] 
-H\sum_{a}m^a \ .
\ee
Taking the limit $n \rightarrow 0$ and using Eqs.  (10) and (12), one obtains
\be
{\cal E}/N=-\frac{cJ}{2}[G_{11}({\bf r}={\bf e})+m^2]-\beta h_0^2G_{con}({\bf r}={\bf 0}) \ ,
\ee
which, owing to Eq. (57), reduces to Eq. (53), as it should be.

Since the interaction between replicas is on-site, the range of the SCOZA direct correlation functions 
$C_{con}^{SCOZA}({\bf r})$ and $C_{dis}^{SCOZA}({\bf r})$ is also unchanged, 
\begin{mathletters}
\be
C_{con}^{SCOZA}({\bf k})=c'_c(m,\tilde J,\tilde h_0)[1-z'(m,\tilde J,\tilde h_0)\hat \lambda({\bf k})]
\ee
\be
C_{dis}^{SCOZA}({\bf k})=c'_d(m,\tilde J,\tilde h_0) \ ,
\ee
\end{mathletters}
so that
\begin{mathletters}
\be
G_{con}^{SCOZA}({\bf r})=\frac{1}{c'_c}P({\bf r},z')
\ee
\be
G_{dis}^{SCOZA}({\bf r})=-\frac{c'_d}{(c'_c)^2}\frac{\partial}{\partial z'}[z'P({\bf r},z')] \ .
\ee
\end{mathletters}
It is then clear that the requirement that Eqs. (13), (50) and (51), together with the exact initial conditions at $\tilde J=0$,
be satisfied implies that  $z'=z$ and 
$c'_c=c_c$ and $c'_d=c_d-c_{01}^2$.

\section{}

We consider the asymmetric bimodal distribution 
\be
{\cal P}(h)=p\delta(h-h_0)+(1-p)\delta(h+h_0)     \ .
\ee
One has ${\bar \tau}=2p-1$ and $c_{00}= [4p(1-p)]^{-1}$. Introducing $f(m,\tilde J,\tilde h_0) =-c_{01}/c_c$ and $r^2(m,\tilde J,\tilde h_0)=(c_{01}^2/c_{00}-c_d)/c_c^2$ 
as new unknown variables, we use Eq. (13) to eliminate $c_c$, i.e., 

\be
c_c=\frac{P(z)}{1-m^2-r^2[P(z)+zP'(z)]}
\ee
and we obtain from Eqs. (50), (51), (56), and (63)  three coupled PDE's in $z(m,\tilde J,\tilde h_0)$, $f(m,\tilde J,\tilde h_0)$ and 
$r(m,\tilde J,\tilde h_0)$,
\begin{mathletters}
\bea
\frac{\partial}{\partial \tilde J}[\frac{(1-z) P(z)}{1-m^2-r^2[P(z)+zP'(z)]}]=&-1&-\frac{1}{2}\frac{\partial ^2}{\partial m^2}
\{ [1-m^2-r^2(P(z)+zP'(z))]\frac{P(z)-1}{zP(z)}\nonumber\\
&+&r^2P'(z)\}
\eea
\be
4p(1-p)\frac{\partial}{\partial \tilde J}[fP(z)]=\frac{1}{2}\frac{\partial}{\partial \tilde h_0}\{ [1-m^2-r^2(P(z)+zP'(z))]\frac{P(z)-1}{zP(z)}+r^2P'(z)\}
\ee
\be
\frac{\partial}{\partial \tilde J}[\frac{fP(z)}{1-m^2-r^2[P(z)+zP'(z)]}]=\frac{1}{4}\frac{\partial ^2}{\partial p \ \partial m}
\{ [1-m^2-r^2(P(z)+zP'(z))]\frac{P(z)-1}{zP(z)}+r^2P'(z)\} \ .
\ee
\end{mathletters}

As explained in section V, the solution of the SCOZA partial differential equations can be easily obtained in terms
of high-temperature series expansions. The remarkable feature of the solution of Eqs. (B3) is that one has
 $c_d/c_c^2\equiv 4p(1-p)f^2-r^2=0 $ through order $\beta^5$. More precisely, for the hypercubic lattice in general dimension, we find

\be
c_d/c_c^2=\frac{4096}{5}d^5\ p^2(1-p)^2\tilde g^2m(1-m^2)^4[-5m+8d\tilde g^{1/2}(p-\frac{1}{2})]\beta^6+O(\beta^7)
\ee
Note that this term cancels out when $m=0$ (this is no longer true for the next-order term). This
implies that  the zero-field susceptibility for the symmetric bimodal distribution is the same through order $\beta ^6$
when calculated from Eqs. (B3) or from Eqs. (66). A more complete study based on a twelve-term series expansion shows 
 that  $c_d=0$ is  a very good approximation when $p=1/2$. On the other hand, it would also be worth studying also the full solution of Eqs. (B3). Indeed,
it is easily seen that  the same equations are obtained for the probability distribution ${\cal P}(h)=
p\delta(h-h_0)+(1-p)\delta(h+h_1)$ (just define $\tau$ by $h=1/2[(h_0+h_1)\tau +(h_0-h_1]$). In mean-field theory, this asymmetric RFIM has 
a rich phase diagram with several coexistence surfaces \cite{MSCCCB91} and it has been claimed to be relevant to
the liquid-vapor coexistence boundary of $^4$He confined in silica aerogel.

\section{}

To illustrate the accuracy of the SCOZA, we show here the terms $a[15]$ in the high-T series expansion 
of the  the zero-field susceptibility, $\chi(m=0)=\sum_{n=0}^{\infty} a[n] (\beta J)^n$. These results can be compared to the exact ones given in 
Ref. \cite{GAAHS93}. One finds

\bea
a[15] &&=32768d^{15}-229376d^{14}+1835008d^{13}/3-3923968d^{12}/5+19566592d^{11}/45 \nonumber\\
& &+62568448d^{10}/
315+2393632768d^9/5775+1309641268736d^8/225225+344496180736d^7/8775 \nonumber\\
& &-250795760924032d^6/259875+1073182266768952832d^5/212837625 \nonumber\\
& &-2633146526329112672d^4/212837625+687607165382389072d^3/42567525 \nonumber\\
& &-6687777748364066392d^2/638512875+522635516549738848d/212837625 \nonumber\\
& &+(-114688d^{13}+
638976d^{12}-1306624d^{11}+3549184d^{10}/3-6071296d^9/15-26430464d^8/45  \nonumber\\
& &-5725256704d^7/10395+530937363712d^6/289575+87605199818176d^5/2027025 \nonumber\\
& &-24080214624448d^4/135135+1584841452347248d^3/6081075-75721982136512d^
2/405405 \nonumber\\
& &+357823952018272d/6081075)g+(184320d^{11}-768000d^{10}+1136640d^9-3053568d
^8/5 \nonumber\\
& &+2969344d^7/15-4845952d^6/105-447235513408d^5/51975+2060640064288d^
4/51975 \nonumber\\
& &-11034060991936d^3/155925+1826181780848d^2/31185-4442599992d/275)g^2 \nonumber\\
& &+(-
547840d^9/3+509952d^8-4678144d^7/9+57472d^6-10716992d^5/315 \nonumber\\
& &+3003435424
d^4/945-8105781856d^3/945+21474530792d^2/2835-4915534784d/1701)g^3 \nonumber\\
& &+(384256d^7/3-187776d^6+1368512d^5/9+2739616d^4/45-85830176d^3/135 \nonumber\\
& &+262363504d^2/315-
10055056d/105)g^4+(-355008d^5/5+52160d^4/3-666224d^3/15 \nonumber\\
& &+23348344d^2/225-
14219344d/225)g^5+(1560464d^3/45+1000808d^2/45-199312d/9)g^6 \nonumber\\
& &-4709644dg^7/315 \nonumber\\
\eea
for the Gaussian distribution function, and
\bea
a[15] & &=32768d^{15} - 229376d^{14}+1835008d^{13}/3 - 3923968d^{12}/5+3940352d^{11}/9+59213824d^{10}/315 \nonumber\\ 
& &+7336961024d^{9}/17325+261515720192d^{8}/45045+26451169582592d^{7}/675675 \nonumber\\ 
& &-1086932402689792d^{6}/1126125+27530806058894912d^{5}/5457375 \nonumber\\ 
& &-877599448749913984d^{4}/70945875+255326320300912d^{3}/15795 \nonumber\\ 
& &-2202231489060446384d^{2}/212837625+1549572126103322224d/638512875 \nonumber\\ 
& &+(-114688d^{13}+638976d^{12}-1306624d^{11}+3528704d^{10}/3-1995776d^9/5 \nonumber\\ 
& &-8554496d^8/15-167141888d^7/297+14174567936d^6/6435+10235242804672d^5/225225 \nonumber\\ 
& &-29834386736192d^4/155925+596984424197072d^3/2027025 \nonumber\\ 
& &-407009452618592d^2/2027025+310791876465968d/6081075)g \nonumber\\ 
& &+(151552d^{11}-1853440d^{10}/3+892928d^9-1246720d^{8}/3+29252864d^{7}/315 \nonumber\\ 
& &-3510912d^{6}/35-65936770624d^{5}/7425+1238778408016d^{4}/31185 \nonumber\\ 
& &-11162124651232d^{3}/155925+7782983079632d^{2}/155925-19615564589086d/467775)g^{2} \nonumber\\ 
& &+(-846848d^{9}/9+3586048d^{8}/15 -6394880d^{7}/27-1122176d^{6}/45-1242688d^{5}/105 \nonumber\\ 
& &+997169632d^{4}/405-52631828384d^{3}/8505+4259397904d^{2}/675 \nonumber\\ 
& &-817609382692d/127575)g^{3}+(1751296d^{7}/63-433024d^{6}/15+1367232d^{5}/35 \nonumber\\ 
& &+115455808d^{4}/4725-1154097376d^{3}/4725+9855809872d^{2}/33075-20089162036d/99225)g^{4} \nonumber\\ 
& &+(-16538048d^{5}/4725-474176d^{4}/567-85810544d^{3}/14175+543224408d^{2}/70875 \nonumber\\ 
& &-3984248d/6075)g^{5}+(65701616d^{3}/467775+18472712d^{2}/155925+23549048d/1403325)g^{6} \nonumber\\ 
& &-35397196dg^{7}/42567525 \nonumber\\ 
\eea
for the bimodal distribution.

\newpage

\begin{center}
\begin{tabular}{|c|cccc|cc|}
\hline
\multicolumn{1}{|c}{}  
&\multicolumn{4}{c|}{$\beta_c J$}&\multicolumn{2}{|c|}{$\gamma$}\\
\hline
\hspace{.25cm} {$\tilde g$} \hspace{2cm}& \hspace{.4cm} RPA\hspace{.25cm}  & \hspace{.25cm} ORPA \hspace{.25cm} 
& \hspace{.25cm} SCOZA \hspace{.25cm} & \hspace{.25cm} ``exact'' \hspace{.25cm}
& \hspace{.25cm} SCOZA \hspace{.25cm} & \hspace{.25cm} ``exact'' \hspace{.25cm}\\
\hline
0 & 0.1000 & 0.1156 & 0.1139 & 0.1139  & 1.031 & 1  \\
0.08 & 0.1087 & 0.1370 & 0.1311 & 0.1315 & 1.094 & 1.12 \\
0.10 & 0.1112 & 0.1438 & 0.1369 & 0.1369 & 1.142 & 1.13   \\
0.12 & 0.1138 & 0.1514 & 0.1425 & 0.1428 & 1.133 & 1.14  \\
0.14 & 0.1165 & 0.1601 & 0.1492 & 0.1493 & 1.148 & 1.142   \\
0.15 & 0.1179 & 0.1649 & 0.1528 & 0.1529 & 1.158 & 1.144 \\
0.18 & 0.1224 & 0.1818 & 0.1656 & 0.1665 & 1.192 & 1.245  \\
0.20 & 0.1257 & 0.1959 & 0.1759 & 0.177 & 1.219 & 1.28 \\
0.25 & 0.1349 & 0.2489 & 0.2090 & 0.213 & 1.190 & 1.30 \\
\hline
\end{tabular}
\end{center}

{Table 1: Inverse critical temperature and critical exponent $\gamma$ for the 5-d hypercubic lattice (Gaussian distribution).
The SCOZA predictions have been obtained from a $[5/6]$ Dlog Pad\'e analysis of the high-T series expansion of the zero-field susceptibility. 
The ``exact'' results are taken from Ref. \cite{GAAHS93}, except for $g=0$ which is taken  from Ref. \cite{A1996}. 
Note that $g$ in Ref. \cite{GAAHS93} corresponds to our $\tilde g$ multiplied by $c^2=100$.}

\newpage

\begin{center}
\begin{tabular}{|c|cccc|cc|}
\hline
\multicolumn{1}{|c}{}  
&\multicolumn{4}{c|}{$\beta_c J$}&\multicolumn{2}{|c|}{$\gamma$}\\
\hline
\hspace{.25cm} {$\tilde g$} \hspace{2cm}& \hspace{.4cm} RPA\hspace{.25cm}  & \hspace{.25cm} ORPA \hspace{.25cm} 
& \hspace{.25cm} SCOZA \hspace{.25cm} & \hspace{.25cm} ``exact'' \hspace{.25cm}
& \hspace{.25cm} SCOZA \hspace{.25cm} & \hspace{.25cm} ``exact'' \hspace{.25cm}\\
\hline
0 & 0.1000 & 0.1156 & 0.1139 & 0.1139 & 1.031 & 1 \\
0.05 & 0.1057 & 0.1296 & 0.1251 & 0.1253 & 1.077 & 1.0\\
0.07 & 0.1085 & 0.1379 & 0.1312 & 0.1314& 1.087& 1.1 \\
0.08 & 0.1100 & 0.1430 & 0.1347 & 0.1349  & 1.090 & 1.11\\
0.09 & 0.1116 & 0.1494 & 0.1386 & 0.1389& 1.086& 1.11 \\
0.10 & 0.1134 & 0.1577 & 0.1431 & 0.1435 &1.077 & 1.135\\
0.11 & 0.1154 & 0.1700 & 0.1473 & 0.1488 &1.005 & 1.15\\
0.12 & 0.1175 & - & 0.1546 & 0.1552 &1.073 & 1.18\\
\hline
\end{tabular}
\end{center}

{Table 2: Same as Table 1 but for the bimodal distribution.}

\large

\vspace*{1.cm}
\begin{figure}[h]
\vspace*{0.5cm}
\hspace*{2.0cm}
\leavevmode
\epsfxsize= 80pt
\epsffile[ 100 370 200 670]{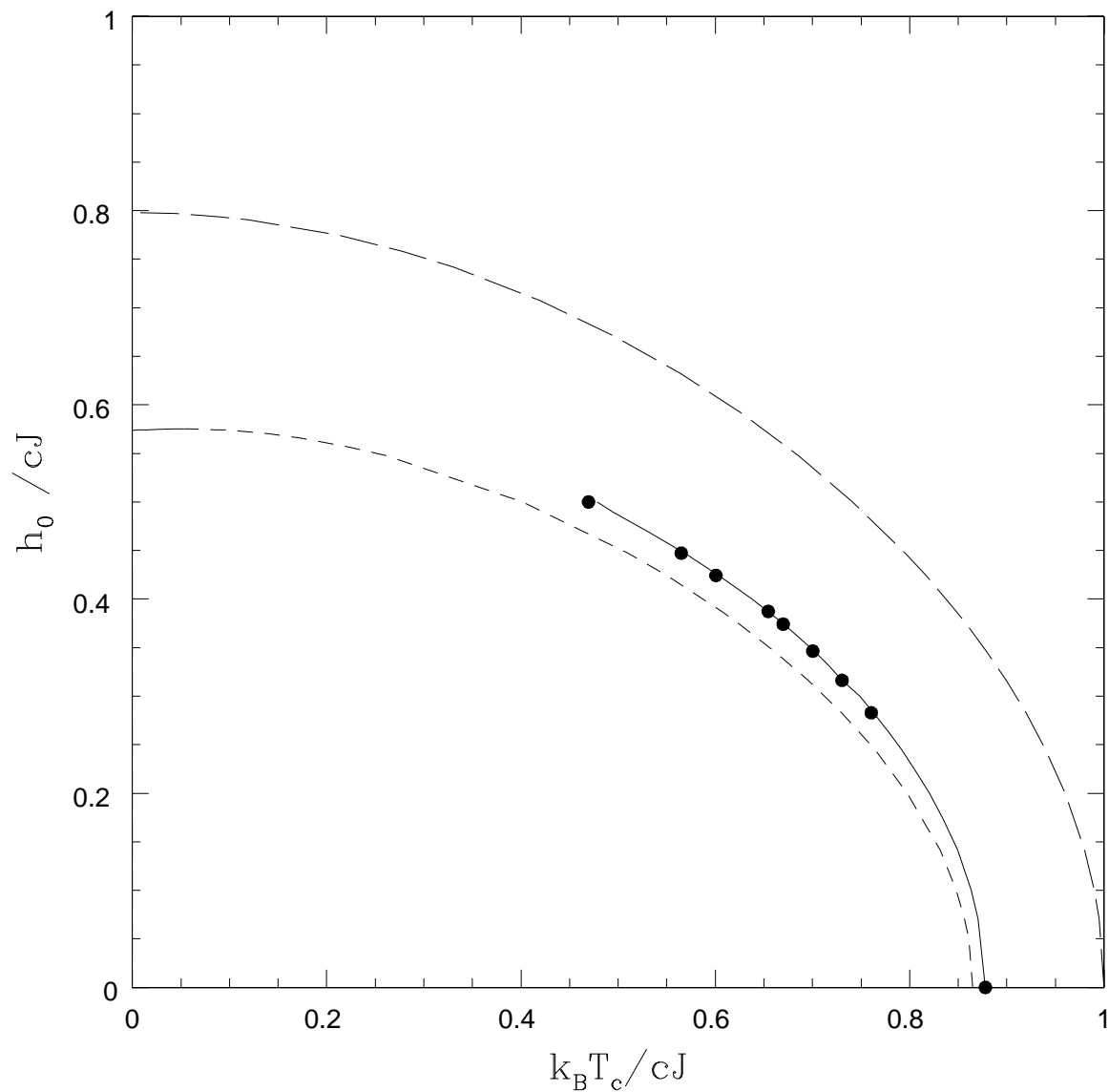}

\vspace*{7.0cm}
\caption{Phase diagram of the random field Ising model with a Gaussian distribution in $d=5$.  Comparison of
the mean-field  (long-dashed line), ORPA  (short-dashed line) and
SCOZA (solid line) predictions to the ``exact'' phase boundary (dots). The SCOZA results are 
obtained from  the high-temperature series expansion of the zero-field
susceptibility. In principle, the SCOZA line should continue down to $T_c=0$ but no reliable  Pad\'e approximant has been 
found at low temperatures.}

\end{figure}

\newpage

\large

\vspace*{1.cm}
\begin{figure}[h]
\vspace*{0.5cm}
\hspace*{2.0cm}
\leavevmode
\epsfxsize= 80pt
\epsffile[ 100 370 200 670]{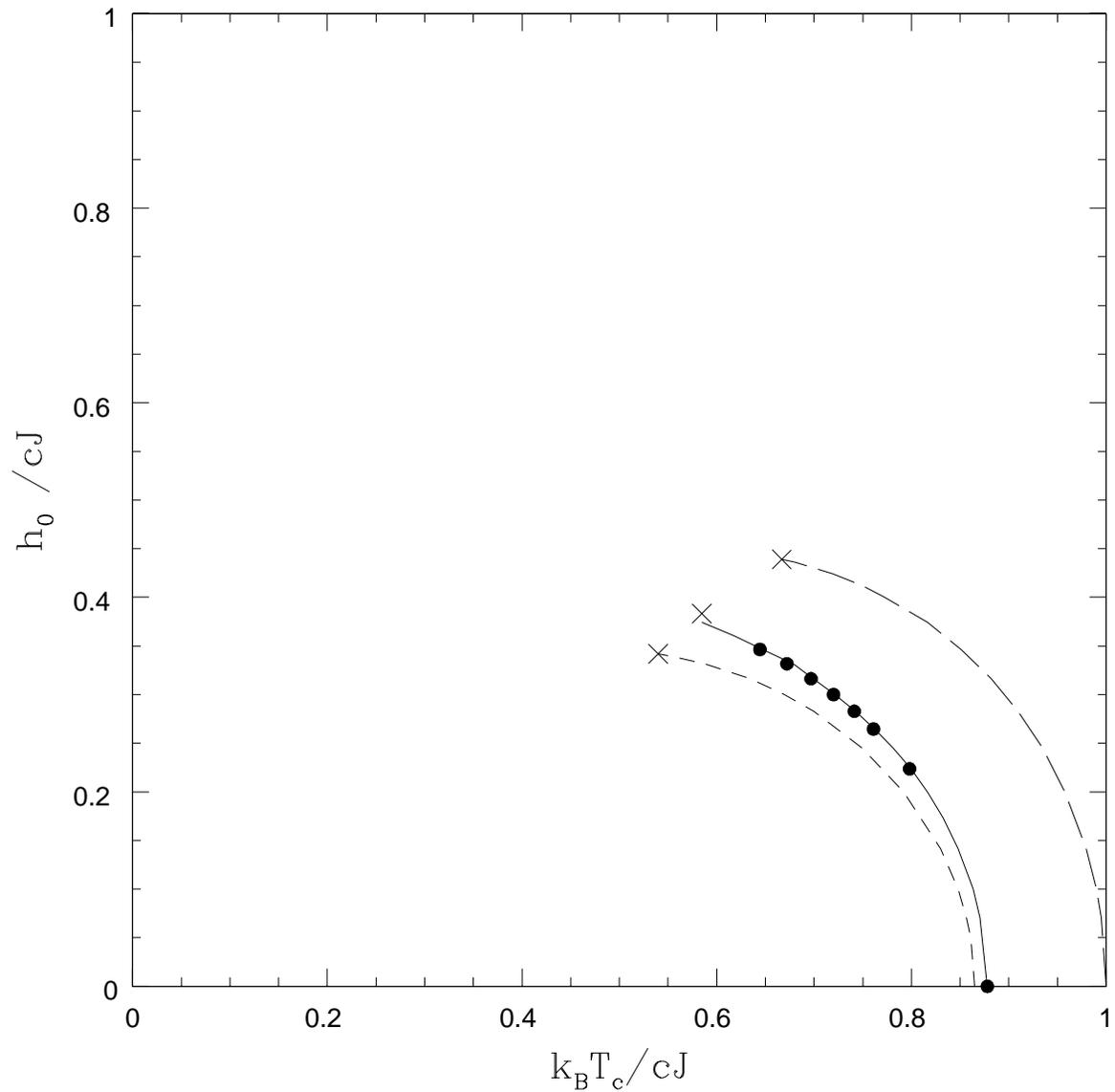}

\vspace*{7.0cm}
\caption{Same as Fig. 1, but for the bimodal distribution. Only the portion of the curves corresponding to a second-order transition
is represented. The crosses locate the tricritical points. The SCOZA tricritical point is estimated from a 1/$d$ series expansion at second order. 
}

\end{figure}
\end{document}